\documentclass[aps,prd,superscriptaddress,nofootinbib,10pt]{revtex4}
\usepackage{amsmath,amssymb}
\usepackage[bookmarks=false,colorlinks=true]{hyperref}
\usepackage{color}
\usepackage{latexsym}
\def\DR{\rm I\kern-1.45pt\rm R}
\def\DC{\kern2pt {\hbox{\sqi I}}\kern-4.2pt\rm C}

\newcommand{\ba}{\begin{array}}
\newcommand{\ea}{\end{array}}
\newcommand{\be}{\begin{equation}}
\newcommand{\ee}{\end{equation}}
\newcommand{\bea}{\begin{eqnarray}}
\newcommand{\eea}{\end{eqnarray}}

\begin{document}
\title{$\mathbb{C}^N$-Smorodinsky-Winternitz system  in a constant magnetic field}
\author{Hovhannes Shmavonyan}
\email{hovhannes.shmavonyan@mail.yerphi.am}
\affiliation{Yerevan Physics Institute, 2 Alikhanian Brothers St., Yerevan  0036 Armenia}

\begin{abstract}
We propose the superintegrable generalization of Smorodinsky-Winternitz system on the $N$-dimensional complex Euclidian space
which is specified by the presence of  constant magnetic field. We find out  that in addition to $2N$ Liouville integrals  the system has additional functionally independent constants of motion, and compute their symmetry algebra.
We perform the Kustaanheimo-Stiefel transformation of $\mathbb{C}^2$- Smorodinsky-Winternitz system to the (three-dimensional) generalized MICZ-Kepler problem and find the symmetry algebra of the latter one. We observe that constant magnetic field appearing in the initial system has no qualitative impact on the resulting system.

\end{abstract}

\maketitle
\section{Introduction}
$N$-dimensional mechanical system will be called {\sl integrable} if it has N  mutually commuting and functionally independent constants of motion. In addition to these constants of motion the system can have additional  ones. In that case we will say that the system is {\sl superintegrable}. Particularly if $N$-dimensional mechanical system has $2N-1$ functionally independent constants of motion it will be called {\sl maximally superintegrable}.
 The one-dimensional singular oscillator is a textbook example of a system which is exactly solvable both on classical and quantum levels.The sum of its $N$ copies, i.e. $N$-dimensional singular isotropic oscillator is, obviously, exactly solvable as well.
It is given by the Hamiltonian
\be\label{SWR}
H=\sum_{i=1}^{N}I_i,\qquad{\rm with}\quad  I_i=\frac{p_i^2}{2}+\frac{g_i^2}{2x_i^2}+\frac{\omega^2x_i^2}{2},\qquad \{p_i,x_j\}=\delta_{ij},\quad \{p_i,p_j\}=\{x_i,x_j\}=0
\ee
  It is not obvious that in addition to Liouville Integrals $I_i$ this system possesses  supplementary series of constants of motion, and is respectively, {\sl maximally superintegrable}, i.e. possesses $2N-1$ functionally independent constants of motion.
All these constants of motion are of the second order on momenta, which guarantee the separability of variables in few  coordinate systems and cause degeneracy of the energy spectrum in quantum level.
 It seems that this  was first noticed  by Smorodinsky and Winternitz, who then investigated superintegrability properties of this system in great detail \cite{sw1,sw2,sw3}. For this reason this model is sometimes called Smorodinsky-Winternitz system and we will use this name as well.  For sure, such a simple and internally rich system would attract wide attention in the community of mathematical and theoretical physics, and that is one of the main reasons why  there are so many publications devoted to its study and further generalizations. Besides the above-mentioned publications, we should as well mention the  references \cite{Evans,evans,evans2,pogosyan,pogosyan2,Hoque,Hoque2,heinzl,miller}(see   the recent  PhD thesis on this subject  with expanded list of references  \cite{PhD}). Notice also  that  Smorodinsky-Winternitz system  is a simplest case of the generalized Calogero model(with oscillator potential) associated with an arbitrary Coxeter root system \cite{cal}. Thus, one hopes that observations done in this simple model could be somehow extended to the Calogero models.
 There is a  well-known superintegrable generalization of the oscillator to sphere, which is known as Higgs oscillator\cite{higgs,higgs2}
 Smorodinsky-Winternitz  model  admits superintegrable generalization  of the  sphere as well \cite{pogosyanSph}.  Though it was first suggested  by Rosochatius in XIX century (without noticing its superintegrability) \cite{rosochatius}, it  was later rediscovered  by many  other authors  as well (e.g. \cite{oksana,oksana2}) . Superintegrable generalization of Calogero model on the sphere also exists \cite{CalogeroSph,CalogeroSph1,CalogeroSph2}.

  In this paper we consider simple generalization of the Smorodinsky-Winternitz system {\sl interacting with constant magnetic field}.
  It is defined on the $N$-dimensional complex Euclidian space parameterized by the coordinates $z^a$ by the Hamiltonian
  \be\label{Hamplan0}
\mathcal{H}=\sum_{a=1}^N\left(\pi_a\bar{\pi}_a+\frac{g_a^2}{z^a\bar{z}^a}+\omega^{2}z^a\bar{z}^a\right),\qquad{\rm with}\quad  \{\pi_a,z^b\}=\delta_{ab},\quad \{{\pi}_a,\bar{\pi}_b\}=\imath B \delta_{ab}
\ee
The (complex) momenta  $\pi_a$  have nonzero Poisson brackets due to the presence of magnetic field  with   magnitude $B$ \cite{SupHam,march}. We will refer this model as $\mathbb{C}^N$-Smorodinsky-Winternitz system. For sure, in the absence of magnetic field this model could be easily reduced to the conventional Smorodinsky-Winternitz model, but the presence of magnetic field could have  nontrivial impact which will be studied in this paper. So, {\sl our main goal is to investigate the whole symmetry algebra of this system}.
Notice that this is not only for academic interest: the matter is that  $\mathbb{C}^1$-Smorodinsky-Winternitz system is a popular model for the qualitative study of the so-called quantum ring \cite{chp,chp1,chp2}, and the study of its behaviour in external magnetic field is quite a natural task. Respectively, $\mathbb{C}^N$-Smorodinsky-Winternitz  could be viewed as an ensemble of $N$ quantum rings interacting with external magnetic field. So  investigation of its symmetry algebra is of the physical importance.

It is well-known for many years that  the energy surface of two-/three-/five-dimensional Coulomb system could be transformed to those of
two-/four-/eight-dimensional oscillator by the use of so-called Levi-Civita-Bohlin/Kustaanheimo-Stiefel/Hurwitz transformation \cite{Bohl,KS,Hur}. More generally,  reducing the two-/four-/eight-dimensional oscillator (-like) models by the action
of $\mathbb{Z}_2/U(1)/SU(2)$ group action, we can get the  two-/three-/five-dimensional Coulomb-like systems specified by the presence of ($\mathbb{Z}_2$/Dirac/Yang)monopole \cite{hopf}. Since $\mathbb{C}^2$-Smorodinsky-Winternitz system is manifestly invariant with respect to $U(1)$ group action, we can  perform its Kustaanheimo-Stiefel transformation, in order to obtain three-dimensional Coulomb-like system. It  was done about ten years ago \cite{mardoyan0}, but in the absence of magnetic field in initial system. Repeating this transformation for the system with constant magnetic field we get unexpected result:  it has no qualitative impact in the resulting system, which was referred in \cite{mardoyan1} as "generalized MICZ-Kepler system"\cite{mic,zw,kep}. In addition, we obtain, in this way, the explicit expression of its symmetry generators and their  symmetry algebra, which as far as we know was not constructed before.

We already mentioned that both oscillator and Smorodinsky-Winternitz system admit superintegrable generalizations to the spheres.
On the other hand the isotropic oscillator on $\mathbb{C}^N$ admits the superintegrable generalization on the complex projective space, moreover, the inclusion of constant magnetic field preserves all symmetries of that system \cite{cpn,cpn2}.
It will be shown that introduction of a constant magnetic field doesn't change these properties of the $\mathbb{C}^N$-Smorodinsky-Winternitz system.
Thus, presented model could be viewed as a  first step towards the construction of the analog of Smorodinsky-Winternitz system on $\mathbb{CP}^N$.

The paper is organized as follows.

In the {\sl Section 2} we review the main properties of the conventional ($\mathbb{R}^N$-)Smorodinsky-Winternitz system, presenting explicit expressions of its symmetry generators, as well as wavefunctions and Energy spectrum.
We also present  symmetry algebra in a very simple, and seemingly new form via  redefinition of  symmetry generators.

In the  {\sl Section 3} we present  $\mathbb{C}^N$-Smorodinsky-Winternitz system in a constant magnetic field, find  the explicit expressions of its  constants of motion. We compute their algebra and find that it is independent from the magnitude of constant magnetic field. Then we quantize a system and obtain wavefunctions and energy spectrum. We notice that the $\mathbb{C}^N$-Smorodinsky-Winternitz system has the same degree of degeneracy as $\mathbb{R}^N$- one, due to the lost part of additional symmetry.
%What is interesting is that the magnetic field  yields the change of the effective frequency only and does not change the form of the spectrum.

In the  {\sl Section 4} we perform Kustaanheimo-Stieffel transformation of the $\mathbb{C}^2$-Smorodinsky-Winternitz system in constant magnetic field and obtain, in this way, the so-called ``generalized MICZ-Kepler system". We find  that  constant magnetic field appearing in the initial system, does not lead to any changes in the resulting one.

In the {\sl Section 5} we discuss the obtained results and possibilites of furthur generalizations. Possible extensions of discussed system include supersymmetrization and quaternionic generalization as well as generalization of  these systems in curved background.

\section{Smorodinsky-Winternitz system on $\mathbb{R}^N$}

Smorodinsky-Winternitz system is defined  as a sum of $N$ copies of one-dimensional singular oscillators \eqref{SWR},
%
%\be
%H=\sum_{i=1}^{N}I_i,
%\quad{\rm
%with}\quad
%I_i=\Big(\frac{p_i^2}{2}+\frac{g_i^2}{2x_i^2}+\frac{\omega^2x_i^2}{2}\Big),\qquad \{p_i,x_j\}=\delta_{ij}.
%\ee
each of them defined by  generators $I_i$ which  obviously form its  Liouville integrals
$\{I_i,I_j\}=0$.
About fifty years ago it was  noticed that this system possesses additional set of constants of motion given by the expressions \cite{sw1}
\be
I_{ij}=L_{ij}L_{ji}-\frac{g_i^2x_j^2}{x_i^2}-\frac{g_j^2x_i^2}{x_j^2},\qquad \{ I_{ij}, H\}=0,
\ee
where $L_{ij}$ are the generators of $SO(N)$ algebra,
\be
 L_{ij}=p_i x_j-p_j x_i\; :\quad \{L_{ij},L_{kl}\}=\delta_{ik}L_{jl}+\delta_{jl}L_{ik}-\delta_{il}L_{jk}-\delta_{jk}L_{il}.
\ee
The generators $I_{ij}$ provides additional $N-1$ functionally independent constants of motions and so this system is maximally superintegrable.
These generators define highly nonlinear symmetry algebra,
\be
\{I_i,I_{jk}\}=\delta_{ij}S_{ik}-\delta_{ik}S_{ij},
\quad
\{I_{ij},I_{kl}\}=\delta_{jk}T_{ijl}+\delta_{ik}T_{jkl}-\delta_{jl}T_{ikl}-\delta_{il}T_{ijk}
\ee
where
\be
S^2_{ij}=-16(I_iI_jI_{ij}+I_i^2g_j^2-I_j^2g_i^2+\frac{\omega^2}{4}I_{ij}^2-g_i^2g_j^2\omega^2), \qquad
T_{ijk}^2=-16(I_{ij}I_{jk}I_{ik}+g_k^2I_{ij}^2+g_j^2I_{ik}^2+g_i^2I_{jk}^2-4g_i^2g_j^2g_k^2).
\ee
The  generators $S^2_{ij}$ and $T^2_{ijk}$ are of the sixth-order in momenta and antisymmetric over $i,j,k$ indices.
The above symmetry algebra could be written in a compact form if we introduce the notation
\be
M_{ij}=I_{ij},\quad M_{0i}=I_i,\quad M_{ii}=g_i^2,\quad M_{00}=\frac{\omega^2}{4}, \quad R_{ijk}=T_{ijk},\quad R_{ij0}=S_{ij}.
\ee
Then one can introduce capital letters   which will take values from $0$ to $N$. It is worth to mention that $M_{IJ}$ is  symmetric, whereas $R_{IJK}$ is antisymmetric with respect to all indices.
In this terms  the whole symmetry algebra of Smorodinsky-Winternitz system  reads
\be
\{M_{IJ},M_{KL}\}=\delta_{JK}R_{IJL}+\delta_{IK}R_{JKL}-\delta_{JL}R_{IKL}-\delta_{IL}R_{IJK}
\ee
where
\be
R_{IJK}^2=-16(M_{IJ}M_{JK}M_{IK}+M_{IJ}^2M_{KK}+M_{IK}^2M_{JJ}+M_{KL}^2M_{II}-4M_{II}M_{JJ}M_{KK})
\ee
One important fact should be mentioned, although in this algebra on the right side we have sum of many terms (square roots), only one term always survives, since in case of three indices are equal, the result is automatically $0$. Consequently in this algebra we always have one square root on the right hand side.
%This system has a superintegrable  analog on a sphere which is known as Rosochatius system.
 Quantum-mechanically the maximal superintegrability is reflected in the dependence of its energy spectrum  on the single,``principal" quantum number only.
Having in mind that in Cartesian coordinates the system decouples to the set of one-dimensional singular oscillators,
we can immediately extract the expressions for its wavefunctions and spectrum from the standard textbooks on quantum mechanics, e.g. \cite{landau},
\be
E_{n|\omega}=\hbar\omega\Big(2n+1+\sum_{i=1}^N\sqrt{\frac14+\frac{g_i^2}{\hbar^2}}\Big),\qquad \Psi=\prod_{i=1}^N\psi(x_i,n_i),
\quad n=\sum_{i=1}^N n_i
\ee
where
\be
\psi(x_{i},n_i)=F\Big(-n_i,1+\sqrt{\frac14+\frac{g_i^2}{\hbar^2}},\frac{\omega x_i^2}{\hbar}\Big)\Big(\frac{\omega x_i^2}{\hbar}\Big)^{\frac{1+\sqrt{1+4g_i^2/\hbar^2}}{4}}e^{-\frac{\omega x_i^2}{2\hbar}}
\ee
Here $F$ is the hypergeometric function.
With these expressions at hands we are ready to study Smorodinsky-Winternitz system on complex Euclidean space in the presence of constant magnetic field.

\section{$\mathbb{C}^N$-Smorodinsky-Winternitz system}
Now let us study $2N$-dimensional  analog of Smorodinsky-Winternitz system interacting with constant magnetic field.
It is defined by \eqref{Hamplan0} and could be viewed as an analog of Smorodinsky-Winternitz system on complex Euclidian space $\left (\mathbb{C}^N, ds^2=\sum_{a=1}^N dz^ad{\bar z}^a \right)$. Thus, we will refer it as $\mathbb{C}^N$-Smorodinsky-Winternitz system.
The analog of SW-system which respects the inclusion of constant magnetic field is defined as follows,
\be\label{Hamplan}
\mathcal{H}=\sum_{a}I_a,\qquad I_{a}=\pi_a\bar{\pi}_a+\frac{g_a^2}{z^a\bar{z}^a}+\omega^{2}z^a\bar{z}^a\;,
\ee
where $z^a, \pi_a$  are complex (phase space) variables with the following non-zero Poisson bracket relations
\be
\{\pi_a,z^b\}=\delta_{ab},\qquad \{\bar{\pi}_a,\bar{z}^b\}=\delta_{ab}, \quad \{{\pi}_a,\bar{\pi}_b\}=\imath B \delta_{ab}.
\ee

For sure, it can be interpreted  as a sum of  $N$ two-dimensional singular oscillators interacting with constant magnetic field perpendicular to the plane. It is obvious that  in addition to $N$ commuting  constants of motion $I_a$
this system has another set of $N$ constants of motion defining
 manifest  $(U(1))^N$ symmetries of the system
\be
L_{a\bar{a}}=\imath(\pi_a z^a-\bar{\pi}_a\bar{z}^a)-{B}z^a{\bar z}^a\; :\{L_{a\bar a}, \mathcal{H}\}=0
%\{L_{a\bar a}, L_{b\bar b}\}=\{L_{a\bar a}, I_b\}=\{I_{a}, I_b\}=0
\ee
and supplementary, non-obvious,   set of constants of motion defined in complete analogy with those of conventional Smorodinsky-Winternitz system:
\be
I_{ab}=L_{a\bar b}L_{b\bar a}+\Big(\frac{g_a^2z^b\bar{z}^b}{z^a\bar{z}^a}+\frac{g_b^2z^a\bar{z}^a}{z^b\bar{z}^b}\Big),\quad
\{I_{ab},\mathcal{H}\}=0,\qquad
 a\neq b
\ee
with $L_{a\bar b}$ being generators of $SU(N)$ algebra
\be\label{SUN}
L_{a\bar b}=\imath(\pi_az^b-\bar{\pi}_b\bar{z}^a)-{B}\bar{z}^a z^b\;:\quad \{L_{a\bar b},L_{c\bar d}\}=i\delta_{a\bar d}L_{c\bar{b}}-i\delta_{c\bar b}L_{c\bar d}.
\ee
These symmetry generators, and $I_a$ obviously commute with  $L_{a\bar a}$ due to manifest $U(1)^N$ symmetry
\be
\{L_{a\bar a}, I_b\}=\{L_{a\bar a}, I_{bc}\}= \{L_{a\bar a}, L_{b\bar b}\}=\{I_{a}, I_b\}=0
\ee
The rest Poisson brackets between them are highly nontrivial
\be
\{I_{a},I_{bc}\}= \delta_{ab}S_{ac}-\delta_{ac}S_{ab},\qquad \{I_{ab},I_{cd}\}=\delta_{bc}T_{abd}+\delta_{ac}T_{bcd}-\delta_{bd}T_{acd}-\delta_{ad}T_{abc} ,
\ee
where
\be
S_{ab}^2=4I_{ab}I_aI_b-(L_{a\bar a}I_b+L_{b \bar b}I_a)^2-4g_a^2I_b^2-4g_b^2I_a^2-4\omega^2I_{ab}(I_{ab}-L_{a\bar a}L_{b\bar b})+4\omega^2g_b^2L_{a\bar a}^2+4g_a^2\omega^2L_{b \bar b}^2+16g_a^2g_b^2\omega^2
\nonumber\ee
\be
-2B(I_{ab}-L_{a\bar a}L_{b \bar b})(L_{a\bar a}I_b+L_{b \bar b}I_a)-B^2(I_{ab}-L_{a \bar a}L_{b \bar b})^2+4B(g_b^2I_aL_{a\bar a}+g_a^2I_bL_{b\bar b})+4B^2g_a^2g_b^2
\ee
\be
T_{abc}^2=2(I_{ab}-L_{a\bar a}L_{b \bar b})(I_{bc}-L_{b\bar b}L_{c \bar c})(I_{ac}-L_{a\bar a}L_{c\bar c})+2I_{ab}I_{ac}I_{bc}+L_{a\bar a}^2L_{b \bar b}^2L_{c\bar c}^2-(I_{bc}^2L_{a \bar a}^2+I_{ab}^2L_{c \bar c}^2+I_{ac}^2L_{b\bar b}^2)
\nonumber\ee
\be
-4(g_c^2I_{ab}(I_{ab} - L_{a\bar a}L_{b\bar b}) + g_a^2I_{bc}(I_{bc} - L_{b \bar b}L_{c \bar c}) +
    g_b^2I_{ac}(I_{ac} - L_{a\bar a}L_{c \bar c})) + 4g_b^2 g_c^2L_{a\bar a}^2 +
 4g_a^2 g_c^2L_{b \bar b}^2 + 4g_a^2 g_b^2L_{c\bar c}^2 + 16g_a^2 g_b^2 g_c^2
\ee
To write the symmetry algebra in a simpler form we can redefine the generators
\be
M_{aa}=L_{a\bar a}^2+4g_a^2,\quad M_{ab}=I_{ab}-\frac{1}{2}L_{a\bar a}L_{b\bar b}, \quad
M_{a0}=I_a-\frac{B}{2}L_{a\bar a},\quad
M_{00}=4\omega^2+B^2.
\ee
Since $L_{a\bar a}$ commute with all other generators Poisson brackets of $M$ will exactly coincide with the Poisson brackets of $I_{ab}$ and $I_a$. Similarly the $R$ tensor is defined as in the real case. So the algebra will have the following form
\be
\{M_{ab},M_{cd}\}=\delta_{bc}T_{abd}+\delta_{ac}T_{bcd}-\delta_{bd}T_{acd}-\delta_{ad}T_{abc} ,\quad  \{M_{a0},M_{ab}\}=\delta_{ab}S_{ac}-\delta_{ac}S_{ab}.
\ee
where
\be
S_{ab}^2=4M_{ab}M_{a0}M_{b0}+\Big(\omega^2+\frac{B^2}{4}\Big)(M_{aa}M_{bb}-4M_{ab}^2)-M_{b0}^2 M_{aa}-M_{a0}^2 M_{bb}
\ee
\be
T_{abc}^2=4M_{ab}M_{bc}M_{ac}-M_{ab}^2M_{cc}-M_{ac}^2M_{bb}-M_{bc}^2M_{aa}+\frac{1}{4}M_{aa}M_{bb}M_{cc}
\ee
Needless to say that $L_{a\bar a}$ commute with all the other constants of motion. Finally
the full symmetry algebra then reads
\be
\{M_{AB},M_{CD}\}=\delta_{BC}R_{ABD}+\delta_{AC}R_{BCD}-\delta_{BD}R_{ACD}-\delta_{AD}R_{ABC}
\ee
where
\be
R_{ABC}^2=4M_{AB}M_{BC}M_{AC}-M_{AB}^2M_{CC}-M_{AC}^2M_{BB}-M_{BC}^2M_{AA}+\frac{1}{4}M_{AA}M_{BB}M_{CC}
\ee
Again capital letters take values from $0$ to $N$. In the complex case $R_{ABC}$ and $M_{AB}$ are again respectively antisymmetric and symmetric as in the real case.
Up to multiplication by a constant this has the same form as the symmetry algebra for the real case.

 Let us briefly discuss the number of conserved quantities. We have $N$ real functionally independent  constants of motion ($I_a$). Moreover  let us mention that $I_{ab}$  is also real, and although it has $N(N-1)/2$ components, the number of functionally independent constants of motion is $N-1$. In addition to this, the complex system has $N$ real conserved quantities  ($L_{a\bar a}$). So the total number of constants of motion is $3N-1$ and it is superintegrable (but not maximally superintegrable). Escpecially if $N=1$ the system is integrable. For $N=2$ the system is superintegrable, but it has only one additional constant of motion. In this case the system is called {\sl minimally superintegrable}.

\subsection*{Quantization}

Quantization will be done using the fact that $\mathbb{C}^N$-Smorodinsky-Winternitz system is a sum of two dimensional singular oscillators. This allows to write the wave function as a product of $N$ wave functions and total energy of the system as a sum of the energies of its subsystems. So the initial problem reduces to two-dimensional one.

\be
\hat I_{a}\Psi_{a}(z_a,\bar z_a)= E_a \Psi_{a}(z_a,\bar z_a) , \quad
\hat{H}\Psi_{tot}=E_{tot}\Psi_{tot}, \quad
\Psi_{tot}=\prod_{a=1}^N\Psi_{a}(z_a,\bar z_{a}), \quad E_{tot}=\sum_{a}^N E_{a}.
\ee
After this reduction, complex indices can be temporarily dropped. Now it is obvious to introduce the momenta operators and commutation relations, which will have the following form in the presence of constant magnetic field.
\be
\hat{\pi}=-\imath(\hbar\partial+\frac{B}{2}\bar{z}),\quad
\hat{\bar\pi}=-\imath(\hbar\bar{\partial}-\frac{B}{2}z)
\quad
[\pi,\bar{\pi}]=\hbar B,\quad [\pi,z]=-\imath\hbar
\ee
Schr{\"o}dinger equation can be written down
\be
\Big[-\hbar^2\partial\bar{\partial}+\Big(\omega^2+\frac{B^2}{4}\Big)z\bar{z}-\hbar\frac{B}{2}(\bar z \bar{\partial}-\partial z)+\frac{g^2}{z\bar z}\Big]\Psi(z,\bar z)=E\Psi(z,\bar z).
\ee
Even in this two-dimensional system additional separation of variables can be done if one writes this system in a polar coordinates using the fact that $z=\frac{r}{\sqrt2}e^{i\phi}$.

\be
\Big[\frac{\partial^2}{\partial r^2}+\frac{1}{r}\frac{\partial}{\partial r}+\frac{2}{\hbar^2}\Big(E+\frac{\hbar^2}{2r^2}\frac{\partial^2}{\partial \phi^2}-\frac{2g^2}{r^2}-\frac12\Big(\omega^2+\frac{B^2}{4}\Big)r^2+\frac{\imath B\hbar}{2}\frac{\partial}{\partial \phi}\Big)\Big]\Psi(r,\phi)=0.
\label{schrodinger}
\ee
Further separation of variables can be done and one can use the fact that
L is a constant of motion.
\be
\Psi(r,\phi)=R(r)\Phi(\phi),\quad \hat{L}\Phi=\hbar m\Phi .
\ee
Using the explicit form of the $U(1)$ generator, normalized solution can be written
\be
\hat L=-\imath\hbar\frac{\partial}{\partial \phi},
\quad
\Phi(\phi)=\frac{1}{\sqrt{2\pi}}e^{\imath m\phi}.
\ee
This result allows to write the equation \eqref{schrodinger} in the following form
\be
\Big[\frac{d^2}{d r^2}+\frac{1}{r}\frac{d}{d r}+\frac{2}{\hbar^2}\Big(E-\frac{\hbar^2 m^2}{2r^2}-\frac{2g^2}{r^2}-\frac12\Big(\omega^2+\frac{B^2}{4}\Big)r^2-\frac{ B\hbar m}{2}\Big)\Big]R(r)=0.
\ee
Solution of this kind of Schr{\"o}dinger equation can be
 %found in \cite{landau} and one can do it by introducing the following notations
%\be
%\zeta=\frac{\sqrt{\omega^2+\frac{B^2}{4}}}{\hbar}r^2,\quad
%m^2+\frac{4g^2}{\hbar^2} =4s^2, \quad
%\frac{E-\frac{Bm}{2}}{\frac{\hbar}{2}\sqrt{\omega^2+\frac{B^2}{4}}}=4(n+s)+2,\quad  P(\zeta)=e^{\zeta/2}\zeta^{-s}R(\zeta)
%\label{notations}
%\ee
%In terms of these variables equation takes a simple form
%\be
%\zeta P''+(2s+1-\zeta)P'+nP=0
%\ee
%Solution of this equation is known and can be written using a hypergeometric function
%\be
%P(\zeta)=F(-n,2s+1,\zeta)
%\ee
%Notations \eqref{notations} allows us to
 written down. The final result for the wave functions of two-dimensional system and the energy spectrum are as follows
\be
\psi(z,\bar z,n,m)=\frac{C_{n,m}}{\sqrt{2\pi}}(\sqrt{z/\bar z})^mF\Big(-n,\sqrt{m^2+\frac{4g^2}{\hbar^2}}+1,\frac{2\sqrt{\omega^2+\frac{B^2}{4}}}{\hbar}z\bar z\Big)\Big(\frac{2\sqrt{\omega^2+\frac{B^2}{4}}}{\hbar}z\bar z\Big)^{1/2\sqrt{m^2+\frac{4g^2}{\hbar^2}}}e^{-\frac{2\sqrt{\omega^2+\frac{B^2}{4}}}{\hbar}z\bar z}
\ee
\be
E=\hbar\sqrt{\omega^2+\frac{B^2}{4}}\Big(2n+1+\sqrt{m^2+\frac{4g^2}{\hbar^2}}\Big)+\frac{B\hbar m}{2}
\ee
Finally the indices of $\mathbb{C}^N$ can be recovered. The total wave function is a product of the wavefunctions and the total energy is the sum of the energies of two-dimensional subsystems
\be
\Psi(z,\bar z)=\prod_{a=1}^N\psi(z_a,\bar z_a,n_a,m_a)
\ee
\be
E_{tot}=\sum_{a=1}^NE_{n_a,m_a}=\hbar\sqrt{\omega^2+\frac{B^2}{4}}\Big(2n+N+\sum_{a=1}^N\sqrt{m_a^2+\frac{4g_a^2}{\hbar^2}}\Big)+\frac{B\hbar }{2}\sum_{a=1}^N m_a,
\ee
\be
n=\sum_{a=1}^N n_a, \quad n=0,1,2... \qquad m_a=0,\pm 1,\pm 2,...
\ee
In contrast to the real case  the energy spectrum of the $\mathbb{C}^N$-Smorodinky-Winternitz system depends on $N+1$ quantum numbers, namely $n$ and $m_a$ .

\section{Kustaanheimo-Stiefel transformation}
There is a well-known  procedure reducing  two-/four-/eight-dimensional oscillator to the two-/three-/five-dimensional Coulomb system.
It is related with the Hopf maps
%$
%S^{2p-1}/S^{p-1}=S^{p}
%$ with $p=1,2,4$
 and assumes the reduction by $\mathbb{Z}_2-/U(1)-/SU(2)-$ group action. In general case it results in the Coulomb like systems specified by the presence of $\mathbb{Z}_2$-/Dirac-/Yang- monopole\cite{hopf,hopf1,hopf2,hopf3}.
Since $\mathbb{C}^N$-Smorodinsky-Winternitz system has manifest $U(1)$ invariance, we could apply its respective reduction procedure related with first Hopf map $S^3/S^1=S^2$, which is known as Kustaanheimo-Stieffel transformation, for the particular case of $N=2$.
Such a reduction was performed  decade ago  \cite{mardoyan0} and was found to be resulted in the so-called
``generalized MICZ-Kepler problem" suggested by  Mardoyan a bit earlier \cite{mardoyan1,mardoyan2}. However the initial system was considered, it was not specified by the presence of constant magnetic field, furthermore, the symmetry algebra of the reduced system was not obtained there. Hence, it is at least deductive to perform Kustaanheimo-Stiefel transformation to the $\mathbb{C}^2$-Smorodinsky-Winternitz  system with constant magnetic field in order to find its impact  (appearing in the initial system) in the resulting one. Furthermore, it is natural way to find  the constants of motion of the ``generalized MICZ-Kepler system" and construct their algebra.

So,  let us perform the reduction of $\mathbb{C}^2$-Smorodinsky-Winternitz system by the $U(1)$-group action given by the generator
\be
J_0=L_{11}+L_{22}=\imath(z\pi-\bar z \bar \pi)-B z\bar z
\ee
For this purpose we have to choose six  independent functions of initial phase space variables which commute with that generators,
\be
q_k=z\sigma_k\bar z,\quad p_k=\frac{z \sigma_k \pi+\bar \pi\sigma_k \bar z }{2z\bar z},\quad k=1,2,3
\ee
where $\sigma_k$ are standard $2\times 2$ Pauli matrices. Matrix indices are dropped here. This transformation is called Kustaanheimo-Stiefel transformation.
Then we calculate their Poisson brackets and fix the value of $U(1)$- generator $J_0=2s$.
As a result, we get the reduced Poisson brackets
\be
\{q_k,q_l\}=0,\quad \{p_k,q_l\}=\delta_{kl},\quad \{p_k,p_l\}=s\epsilon_{klm}\frac{q_m}{|q|^3}
\ee
Expressing the Hamiltonian via $q_i, p_i, J_0$ and fixing the value of the latter one, we get
\be
H_{SW}=2|q|\Big[\frac{p^2}{2}+\frac{s^2}{2|q|^2}+\frac{Bs}{2|q|}+\frac{1}{2}\Big(\frac{B^2}{4}+\omega^2\Big)+\frac{g_1^2}{|q|(|q|+q_3)}+\frac{g_2^2}{|q|(|q|-q_3)}\Big]
\ee
So, we reduced the $\mathbb{C}^2$-Smorodinsky-Winternitz Hamiltonian to the three-dimensional system.
To get the Coulomb-like system we fix the  energy surface or reduced Hamiltonian, $H_{SW}-E_{SW}=0$ and divide it on $2|q|$.
This yields the equation
\be
{\cal H}_{gMICZ}-\mathcal{E}=0,\qquad {\rm with}\quad \mathcal{E}\equiv -\frac{\omega^2+{B^2}/{4} }{2}
\ee
and
\be
{\cal H}_{gMICZ}=\frac{p^2}{2}+\frac{s^2}{2|q|^2}+\frac{g_1^2}{|q|(|q|+q_3)}+\frac{g_2^2}{|q|(|q|-q_3)}-\frac{\gamma}{|q|} \qquad {\rm with}\quad \gamma\equiv\frac{E_{SW}-Bs}{2}.
\ee
The latter expression defines the Hamiltonian of ``generalized MICZ-Kepler problem". Hence, we transformed the energy surface of the reduced $\mathbb{C}^2$-Smorodinsky-Winternitz Hamiltonian to those of (three-dimensional) ``Generalized MICZ-Kepler system".
Additionally it has an inverse square potential and this system has an interaction with a Dirac monopole magnetic field which affects the symplectic structure.

{\sl Surprisingly, the reduced system contains interaction with Dirac monopole field only, i.e. the constant magnetic field in the original system does not contribute in the reduced one. All dependence on $B$ is hidden in $s$ and $\gamma$, which are fixed,  so the reduced system does not depend on $B$ explicitly.}

Now this reduction can be done for constants of motion. Before doing that it is convenient to present the
initial generators of $u(2)$ algebra  given by \eqref{SUN} in the form
\be
J_{0}=i(z\pi-\bar z \bar \pi)-{Bz\bar z}, \quad J_k=\frac{i}{2}(z\sigma_k\pi- \bar \pi \sigma_k\bar z)-\frac{Bz\sigma_k\bar z}{2}\; :\{J_0, J_i\}=0,\quad \{J_i,J_j\}=\varepsilon_{ijk}J_k.
\ee
After reduction  we get   $J_{0}
=2s$.   After the reduction, the rest $su(2)$ generators result in the  generators of the $so(3)$ rotations of three-dimensional Euclidian space
with  the Dirac monopole placed in the beginning of Cartesian coordinate frame,
\be
J_k=\epsilon_{klm}p_{l}q_{m}-s\frac{q_k}{|q|}
\ee
Then the symmetry generators for the ``generalized MICZ-Kepler system" can be written down,
\be
{\cal I}=\frac{I_1-I_2}{2}+\frac{B}{4}(L_{22}-L_{11})=p_1J_2-p_2J_1+\frac{x_3\gamma}{r}+\frac{g_1^2(r-x_3)}{r(r+x_{3})}-\frac{g_2^2(r+x_3)}{r(r-x_3)}
\ee
\be
{\cal L}=\frac{1}{2}(L_{22}-L_{11})=J_{3}=p_{1}q_{2}-q_{1}p_{2}-\frac{sq_{3}}{|q|},\qquad
{\cal J}=I_{12}=J_1^2+J_2^2+\frac{g_1^2(r-q_3)}{r+q_3}+\frac{g_2^2(r+q_3)}{r-q_3}.
\ee
It is important to notice that ${\cal I}$ is a generalization of the $z$-component of the Runge-Lenz vector.

The  relation of the initial system and the reduced one will allow to find the symmetry algebra of the final system
  using the previously obtained result for the complex Smorodinsky-Winternitz system. First of all the constants of motion in the initial system will also commute with the reduced Hamiltonian.
\be
\{{\cal H}_{gMICZ},{\cal I}\}=\{{\cal H}_{gMICZ},{\cal J}\}=\{{\cal H}_{gMICZ},{\cal L}\}=0
\ee
Moreover, since in the initial system $L_{a\bar a}$ generators commute with all the other constants of motion one can write.
\be
\{{\cal L},{\cal J}\}=\{{\cal L},{\cal I}\}=0
\ee
There is only one non-trivial commutator
\be
\{{\cal I},{\cal J}\}=S
\ee
$S$ here coincides with $S_{12}$ of $\mathbb{C}^{2}$-Smorodinsky-Winternitz system and can be written using the generators of the reduced system.
\be
S^2=2{\cal H}_{gMICZ}\Big[4\Big({\cal J}+\frac{1}{2}\Big({\cal L}^2-s^2\Big)\Big)^2-\Big(4g_2^2+({\cal L}+s)^2\Big)\Big(4g_1^2+({\cal L}-s)^2\Big)\Big]
-\Big(4g_2^2+({\cal L}+s)^2\Big)\Big({\cal I}+\gamma\Big)^2
\nonumber
\ee
\be
-\Big(4g_1^2+({\cal L}-s)^2\Big)\Big({\cal I}-\gamma\Big)^2-4\Big({\cal J}+\frac{1}{2}({\cal L}^2-s^2)\Big)\Big({\cal I}-\gamma\Big)\Big({\cal I}+\gamma\Big)
\ee
There is a crucial fact that should be mentioned. Although the initial system had an interaction with magnetic field, after reduction we don't have any dependence on $B$ both in symplectic structure and in generators of the symmetry algebra, at least in classical level. In other words, the reduced system does not feel the  magnetic field of the initial system.

%\subsection{Supersymmetrization}
%The Hamiltonian \eqref{Hamplan} can be represented in the form
%\be
%\mathcal{H}=\sum_{a=1}^N\left(\pi_a\bar\pi_a + %\omega^2\partial_{a}K \partial_{\bar a}K %\right)=\sum_{a}\left(\pi_a\bar{\pi}_a+\frac{g_a{\bar %g}}{z^a\bar{z}^a}+\omega^{2}z^a\bar{z}^a+g_a+{\bar g}_a\right)
%,
%\ee
%where
%\be
%K= z\bar z +\frac{1}{\omega}\sum_{a=1}^N (g_a\log z^a +{\bar %g}_a\log{\bar z}^a )
%\ee
%is kahler potential on $\mathbb{C}^N$.
%Hence, it admits Weak $\mathcal{N}=4$ supersymmetry

\section{Discussion and outlook}

%We have written down explicit form of the symmetry algebra of real Smorodinsky-Winternitz system.
In this paper we formulated  the analog of the Smorodinksy-Winternitz system  interacting with a constant magnetic field on the $N$-dimensional complex Euclidian space $\mathbb{C}^N$. We found out it has $3N-1$ functionally independent  constants of motion and derived
the symmetry algebra of this system.
Quantization of these systems is also discussed. While  for the real Smorodinsky-Winternitz system energy spectrum is totally degenerate and depends on single ("principal") quantum number, the  $\mathbb{C}^N$-Smorodinsky-Winternitz  energy spectrum depends on $N+1$ quantum numbers. Then we performed Kustaanheimo-Stiefel transformation of the $\mathbb{C}^2$-Smorodinsky-Winternitz system and reduced it to the  so-called "generalized  MICZ-Kepler problem". We   obtained the symmetry algebra of the latter system using the result obtained for the initial ones. Moreover, we have shown that the presence of constant magnetic field in the initial problem does not affect the reduced system.

There are several generalizations one can perform for this system. Straightforward   task is the construction of a quaternionic ($\mathbb{H}^N$-)  analog of this system. While complex structure allows to introduce constant magnetic field without violating the superintegrability, quaternionic structure should  allow to introduce interaction with  $SU(2)$ instanton.
It seems that one can also introduce the superintegrable analogs of the $\mathbb{C}^N$-/$\mathbb{H}^N$-Smorodinsky-Winternitz  systems on
the complex/quaternionic   projective space $\mathbb{CP}^N$/$\mathbb{HP}^N$, having in mind the existence of such generalization for the $\mathbb{C}^N$-/($\mathbb{H}^N$-) oscillator \cite{cpn,hpn}. We expect that the  inclusion of a constant magnetic/instanton field does not cause any qualitative changes for this system.
 These generalizations will be discussed later on.
 %
%Finally,  it is worth to mention that $\mathbb{C}^N$-Smorodinsky-Winternitz system has the form of a so called K\"ahler oscillator and there is a universal procedure to perform $N=4$ "weak" supersymmetrization of this system \cite{kahler1,kahler2}.

\acknowledgements
I am grateful to  Armen Nersessian for his impact on this work, which includes suggestion of the problem and introduction to subject, numerous discussions and help in preparation of manuscript. Thanks to his great support and encouragement this work is  finally completed.
I am also thankful to Lusine Goroyan and Ruben Hasratyan for editing.

This work was done within   ICTP Affiliated Center programs   AF-04 and Regional Doctoral Program on Theoretical and Experimental Particle Physics sponsored by Volkswagen Foundation. I also acknowledge  partial financial support within  research grant from the ANSEF-Armenian National Science and Education Fund based in New York, USA.

\end{document}